\documentclass{article}
\usepackage{amsfonts}

\usepackage{graphicx}
\usepackage{amsmath}
\usepackage{a4wide}

\newtheorem{example}{Example}

\makeatletter
\def\@removefromreset#1#2{\let\@tempb\@elt
     \def\@tempa#1{@&#1}\expandafter\let\csname @*#1*\endcsname\@tempa
     \def\@elt##1{\expandafter\ifx\csname @*##1*\endcsname\@tempa\else
    \noexpand\@elt{##1}\fi}     \expandafter\edef\csname cl@#2\endcsname{\csname cl@#2\endcsname}     \let\@elt\@tempb
     \expandafter\let\csname @*#1*\endcsname\@undefined}

\@removefromreset{equation}{section}

\@removefromreset{theorem}{section}
\makeatother
\input{tcilatex}

\begin{document}

\title{On separability of quantum states and the violation of Bell-type
inequalities }
\author{Elena R. Loubenets\thanks{%
This work is partially supported by MaPhySto - A Network in Mathematical
Physics and Stochastics, funded by The Danish National Research Foundation.} 
\\
Department of Mathematical Sciences\\
University of Aarhus}
\maketitle

\begin{abstract}
In contrast to the wide-spread opinion that any separable quantum state
satisfies every classical probabilistic constraint, we present a simple
example where a separable quantum state does not satisfy the original Bell
inequality although the latter inequality, in its perfect correlation form,
is valid for all joint classical measurements.

In a very general setting, we discuss inequalities for joint experiments
upon a bipartite quantum system in a separable state. We derive quantum
analogues of the original Bell inequality and specify the conditions
sufficient for a separable state to satisfy the original Bell inequality. We
introduce the extended CHSH inequality and prove that, for any separable
quantum state, this inequality holds for a variety of linear combinations.
\end{abstract}

\section{Introduction}

The relation between non-separability and the violation of Bell-type
inequalities is discussed in many papers (see, for example, the review [1]
and references therein). It has been argued that any separable quantum state
satisfies every Bell-type inequality.

A Bell-type inequality is usually viewed as any constraint on averages or
probabilities arising under the description of joint experiments in the
classical probabilistic frame. In this paper, we consider a simple example
where, for a joint experiment upon the two-qubit system in a separable
state, the original Bell inequality [2-4], in its perfect correlation form,
is violated. The latter inequality is, however, valid for all joint
classical measurements\footnote{%
See appendix.}.

In a very general setting, we analyze inequalities arising for quantum
''locally realistic'' [5] joint experiments on a bipartite quantum system in
a separable state. We derive quantum analogues of the original Bell
inequality and specify the conditions where a separable state satisfies the
original Bell inequality, in its perfect correlation or anti-correlation
forms. These sufficient conditions include Bell's correlation restrictions
as particular cases.

We introduce the extended CHSH inequality for a linear combination of mean
values and prove that, for any separable quantum state, this inequality is
valid for a variety of linear combinations. The latter fact may be used for
distinguishing between separable and non-separable quantum states via
computer processing of linear combinations of statistical averages.

\section{Violation of Bell's inequality for a separable state}

Consider a bipartite quantum system $\mathcal{S}_{q}+\mathcal{S}_{q}$,
described in terms of the tensor product $\mathbb{C}^{2}\otimes \mathbb{C}%
^{2}.$

Following the presentation in [6], page 156, denote by 
\begin{equation}
J^{(\theta )}=\left\{ |\uparrow \rangle \langle \uparrow |-|\downarrow
\rangle \langle \downarrow |\right\} cos2\theta +\left\{ |\uparrow \rangle
\langle \downarrow |+|\downarrow \rangle \langle \uparrow |\right\}
sin2\theta  \label{op}
\end{equation}
a self-adjoint operator on $\mathbb{C}^{2}$ with eigenvalues $\lambda
_{1,2}^{(\theta )}=\pm 1$. Here, $\theta $ is a real valued parameter
representing from the physical point of view an angle from some axis (say, $%
z $-axis) and by the symbols $|\uparrow \rangle $ and $|\downarrow \rangle $
we denote eigenvectors of the operator $J^{(0)},$ corresponding to the
eigenvalues $(+1)$ and $(-1),$ respectively. Formally, the operator $%
J^{(\theta )}$ represents the spin operator $S_{z}cos2\theta
+S_{x}sin2\theta $.

Under a joint experiment, constituting a measurement of a quantum observable 
$J^{(\theta _{1})}\otimes J^{(\theta _{2})},$ $\forall \theta _{1},\theta
_{2},$ and performed upon the two-qubit system in a state $\rho ,$ the
quantum average 
\begin{equation}
\langle J^{(\theta _{1})}\otimes J^{(\theta _{2})}\rangle _{\rho }:=\mathrm{%
tr}[\rho \{J^{(\theta _{1})}\otimes J^{(\theta _{2})}\}],  \label{2'}
\end{equation}
represents the expectation value 
\begin{equation}
\langle \lambda ^{(\theta _{1})}\lambda ^{(\theta _{2})}\rangle _{\rho
}=\langle J^{(\theta _{1})}\otimes J^{(\theta _{2})}\rangle _{\rho }
\label{vg}
\end{equation}
of the product $\lambda ^{(\theta _{1})}\lambda ^{(\theta _{2})}$ of the
observed outcomes.

For three different joint experiments upon $\mathcal{S}_{q}+\mathcal{S}_{q}$%
, constituting measurements of quantum observables 
\begin{equation}
J^{(\theta _{a})}\otimes J^{(\theta _{b})},\text{ \ \ \ \ }J^{(\theta
_{a})}\otimes J^{(\theta _{c})}\text{, \ \ \ \ }J^{(\theta _{b})}\otimes
J^{(\theta _{c})},
\end{equation}
respectively, consider the corresponding expectation values 
\begin{equation}
\langle \lambda ^{(\theta _{a})}\lambda ^{(\theta _{b})}\rangle _{\rho },%
\text{ \ \ \ }\langle \lambda ^{(\theta _{a})}\lambda ^{(\theta
_{c})}\rangle _{\rho },\text{ \ \ \ }\langle \lambda ^{(\theta _{b})}\lambda
^{(\theta _{c})}\rangle _{\rho },  \label{up}
\end{equation}
for the case where the two-qubit is initially in the same state $\rho $.
(Here, $a,b,c$ are indices, specifying different $\theta $)$.$

Take a separable initial quantum state\footnote{%
Since quantum sub-systems are identical, we take a symmetrized initial
density operator.} 
\begin{equation}
\rho _{0}=\frac{1}{2}\left\{ |\uparrow ><\uparrow |\otimes |\downarrow
><\downarrow |+|\downarrow ><\downarrow |\otimes |\uparrow ><\uparrow
|\right\} .  \label{state}
\end{equation}
Then the corresponding expectation values (\ref{up}) are given by: 
\begin{eqnarray}
\langle \lambda ^{(\theta _{a})}\lambda ^{(\theta _{b})}\rangle _{\rho _{0}}
&=&\mathrm{tr}[\rho _{0}\{J^{(\theta _{a})}\otimes J^{(\theta
_{b})}\}]=-cos2\theta _{a}\text{ }cos2\theta _{b},\text{ }  \label{gy} \\
\langle \lambda ^{(\theta _{a})}\lambda ^{(\theta _{c})}\rangle _{\rho _{0}}
&=&\mathrm{tr}[\rho _{0}\{J^{(\theta _{a})}\otimes J^{(\theta
_{c})}\}]=-cos2\theta _{a}\text{ }cos2\theta _{c},  \notag \\
\langle \lambda ^{(\theta _{b})}\lambda ^{(\theta _{c})}\rangle _{\rho _{0}}
&=&\mathrm{tr}[\rho _{0}\{J^{(\theta _{b})}\otimes J^{(\theta
_{c})}\}]=-cos2\theta _{b}\text{ }cos2\theta _{c},  \notag
\end{eqnarray}

For the considered three joint experiments, the original Bell inequality, in
its perfect correlation form\footnote{%
See [2-4] and also [6], page 163, and the formula (\ref{34}) in appendix.}, 
\begin{equation}
|\langle \lambda ^{(\theta _{a})}\lambda ^{(\theta _{b})}\rangle _{\rho
_{0}}-\langle \lambda ^{(\theta _{a})}\lambda ^{(\theta _{c})}\rangle _{\rho
_{0}}|\leq 1-\langle \lambda ^{(\theta _{b})}\lambda ^{(\theta _{c})}\rangle
_{\rho _{0}}  \label{Berrlor}
\end{equation}
reads 
\begin{equation}
|cos2\theta _{a}cos2\theta _{b}-cos2\theta _{a}cos2\theta _{c})|\leq
1+cos2\theta _{b}cos2\theta _{c}.  \label{bell'}
\end{equation}

Let, for example, $\theta _{a}=0,$ $\theta _{b}=\pi /6,$ $\theta _{c}=\pi
/3, $ then 
\begin{equation}
cos2\theta _{a}=1,\text{ \ \ \ \ }cos2\theta _{b}=1/2,\text{ \ \ \ \ }%
cos2\theta _{c}=-1/2.  \label{kk}
\end{equation}
Substituting (\ref{kk}) into (\ref{bell'}), we derive the obvious violation.

It is also easy to verify the violation of (\ref{Berrlor}) if the
expectation values, standing in this inequality, are expressed via the
symmetrized tensor products (for details, see section 3.2): 
\begin{eqnarray}
\langle \lambda ^{(\theta _{a})}\lambda ^{(\theta _{b})}\rangle _{\rho _{0}}
&=&\mathrm{tr}[\rho _{0}\{J^{(\theta _{a})}\otimes J^{(\theta
_{b})}\}_{sym}], \\
\langle \lambda ^{(\theta _{a})}\lambda ^{(\theta _{c})}\rangle _{\rho _{0}}
&=&\mathrm{tr}[\rho _{0}\{J^{(\theta _{a})}\otimes J^{(\theta
_{c})}\}_{sym}],  \notag \\
\langle \lambda ^{(\theta _{b})}\lambda ^{(\theta _{c})}\rangle _{\rho _{0}}
&=&\mathrm{tr}[\rho _{0}\{J^{(\theta _{b})}\otimes J^{(\theta
_{c})}\}_{sym}],  \notag
\end{eqnarray}
where we introduce the notation 
\begin{equation}
\{V_{1}\otimes V_{2}\}_{sym}:=\frac{1}{2}\{V_{1}\otimes V_{2}+V_{2}\otimes
V_{1}\},  \label{def}
\end{equation}
for any operators $V_{1},V_{2}$.

\emph{The above simple example shows that the statement that\ any separable
state satisfies every classical constraint is false.}\bigskip

For the considered example, let also check the validity of the original CHSH
inequality [7].

Suppose that we have four different joint experiments, constituting
measurements of quantum observables 
\begin{equation}
J^{(\theta _{a})}\otimes J^{(\theta _{b})},\text{ \ \ }J^{(\theta
_{c})}\otimes J^{(\theta _{b})},\text{ \ \ }J^{(\theta _{c})}\otimes
J^{(\theta _{d})}\text{ \ \ }J^{(\theta _{a})}\otimes J^{(\theta _{d})},
\end{equation}
and performed on the two-qubit in a state $\rho _{0}$. Here the indices $%
a,b,c,d$ specify different $\theta .$

For these four joint experiments, consider the left-hand side of the
original CHSH inequality: 
\begin{equation}
\left| \langle \lambda ^{(\theta _{a})}\lambda ^{(\theta _{b})}\rangle
_{\rho _{0}}+\langle \lambda ^{(\theta _{c})}\lambda ^{(\theta _{b})}\rangle
_{\rho _{0}}+\langle \lambda ^{(\theta _{c})}\lambda ^{(\theta _{d})}\rangle
_{\rho _{0}}-\langle \lambda ^{(\theta _{a})}\lambda ^{(\theta _{d})}\rangle
_{\rho _{0}}\right| \leq 2.  \label{CHSH}
\end{equation}
From (\ref{gy}) it follows, 
\begin{eqnarray}
&&\left| \langle \lambda ^{(\theta _{a})}\lambda ^{(\theta _{b})}\rangle
_{\rho _{0}}+\langle \lambda ^{(\theta _{c})}\lambda ^{(\theta _{b})}\rangle
_{\rho _{0}}+\langle \lambda ^{(\theta _{c})}\lambda ^{(\theta _{d})}\rangle
_{\rho _{0}}-\langle \lambda ^{(\theta _{a})}\lambda ^{(\theta _{d})}\rangle
_{\rho _{0}}\right|  \label{dr'} \\
&=&\left| \text{ }cos2\theta _{a}cos2\theta _{b}+cos2\theta _{c}cos2\theta
_{b}+cos2\theta _{c}cos2\theta _{d}-cos2\theta _{a}cos2\theta _{d}\right| . 
\notag
\end{eqnarray}
Due to the inequality 
\begin{equation}
|x-y|\leq 1-xy,  \label{10'}
\end{equation}
which is valid, for any $|x|\leq 1,$ $|y|\leq 1,$ we further derive for the
right-hand side of (\ref{dr'}): 
\begin{eqnarray}
&&\left| \text{ }cos2\theta _{a}cos2\theta _{b}+cos2\theta _{c}cos2\theta
_{b}+cos2\theta _{c}cos2\theta _{d}-cos2\theta _{a}cos2\theta _{d}\right| \\
&\leq &\left| \text{ }cos2\theta _{a}cos2\theta _{b}+cos2\theta
_{c}cos2\theta _{b}\right| +\left| \text{ }cos2\theta _{c}cos2\theta
_{d}-cos2\theta _{a}cos2\theta _{d}\right|  \notag \\
&\leq &1+cos2\theta _{a}cos2\theta _{c}+1-cos2\theta _{a}cos2\theta _{c} 
\notag \\
&\leq &2.  \notag
\end{eqnarray}
\medskip

Hence, \emph{in the considered example, the original CHSH inequality holds
true} \emph{although the original Bell inequality, in its perfect
correlation form, is violated.}

Both, the original CHSH inequality and the original Bell inequality (in its
perfect correlation form), represent the constrains valid under all joint
classical measurements.\emph{\ }

\section{Quantum inequalities for a separable state}

In a very general setting, let analyze inequalities arising for quantum
''locally realistic'' (see [5]) joint experiments of the Alice/Bob type,
performed on a bipartite quantum system in a separable state.

\subsection{General case}

Let $\mathcal{S}_{q}^{(1)}\mathcal{+S}_{q}^{(2)}$ be a bipartite quantum
system, described in terms of a separable complex Hilbert space $\mathcal{H}%
_{1}\otimes \mathcal{H}_{2}.$

Consider two joint experiments on $\mathcal{S}_{q}^{(1)}\mathcal{+S}%
_{q}^{(2)}$, each with real valued outcomes in a set $\Lambda _{1}$ and in a
set $\Lambda _{2}$, and described by the POV measures: 
\begin{align}
M^{(\alpha ,\beta _{1})}(B_{1}\times B_{2})& =M_{1}^{(\alpha
)}(B_{1})\otimes M_{2}^{(\beta _{1})}(B_{2}),  \label{pov} \\
M^{(\alpha ,\beta _{2})}(B_{1}\times B_{2})& =M_{1}^{(\alpha
)}(B_{1})\otimes M_{2}^{(\beta _{2})}(B_{2}),  \notag
\end{align}
for any subset $B_{1}$ of $\Lambda _{1}$ and any subset $B_{2}$ of $\Lambda
_{2}.$

In (\ref{pov}), the parameters $\alpha ,\beta \in \Gamma $ are of any
nature\ and characterize set-ups of the corresponding (marginal) experiments
with outcomes in $\Lambda _{1}$ and in $\Lambda _{2},$ respectively. ($%
\Lambda _{1}$ may be thought as a set of outcomes on the ''side'' of Alice
and $\Lambda _{2}$ as a set of outcomes on the ''side'' of Bob).

For simplicity, we suppose that the absolute value of each observed outcome
is bounded, that is: 
\begin{equation}
\Lambda _{1}=\{\lambda _{1}\in \mathbb{R}:|\lambda _{1}|\leq \mathrm{C}%
_{1}\},\text{ \ \ \ \ \ }\ \Lambda _{2}=\{\lambda _{2}\in \mathbb{R}%
:|\lambda _{2}|\leq \mathrm{C}_{2}\},  \label{hu}
\end{equation}
with some $C_{1}>0,$ $C_{2}>0.$

For each of the joint experiments (\ref{pov}), performed on\ a bipartite
quantum system $\mathcal{S}_{q}^{(1)}\mathcal{+S}_{q}^{(2)}$ in an initial
state $\rho :$

\begin{itemize}
\item  the formula 
\begin{equation}
\mathrm{tr}[\rho \{M_{1}^{(\alpha )}(B_{1})\otimes M_{2}^{(\beta
_{n})}(B_{2})\}],\text{ \ \ \ \ \ \ }n=1,2,
\end{equation}
represents the joint probability of an outcome $\lambda _{1}$ to belong to a
subset $B_{1}\subseteq $ $\Lambda _{1}$ and an outcome $\lambda _{2}$ to
belong to a subset $B_{2}\subseteq \Lambda _{2}$, that is, of a compound
outcome $(\lambda _{1},\lambda _{2})\in B_{1}\times B_{2};$

\item  the formula 
\begin{align}
\langle \lambda _{1}\lambda _{2}\rangle _{\rho }^{(\alpha ,\beta _{n})}&
:=\int_{\Lambda _{1}\times \Lambda _{2}}\lambda _{1}\lambda _{2}\mathrm{tr}%
[\rho \{M_{1}^{(\alpha )}(d\lambda _{1})\otimes M_{2}^{(\beta n)}(d\lambda
_{2})\}]  \label{sata.aver} \\
& =\mathrm{tr}[\rho (W_{1}^{(\alpha )}\otimes W_{2}^{(\beta _{n})})],\text{
\ \ \ }n=1,2,  \notag
\end{align}
represents the expectation value of the product $\lambda _{1}\lambda _{2}$
of the observed outcomes.

Here, by $W_{1}^{(\alpha )}$ and $W_{2}^{(\beta _{i})},$ we denote the
self-adjoint bounded linear operators on $\mathcal{H}_{1}$ and $\mathcal{H}%
_{2},$ respectively, defined by the relations 
\begin{align}
W_{1}^{(\alpha )}& =\int_{\mathbb{\Lambda }_{1}}\lambda _{1}M_{1}^{(\alpha
)}(d\lambda _{1}),\text{ \ \ \ \ \ \ \ \ \ \ }\left\| W_{1}^{(\alpha
)}\right\| \leq \mathrm{C}_{1}  \label{operat} \\
W_{2}^{(\beta _{n})}& =\int_{\mathbb{\Lambda }_{2}}\lambda _{2}M_{2}^{(\beta
_{n})}(d\lambda _{2})\text{, \ \ \ }\ \ \ \ \ \ \left\| W_{2}^{(\beta
_{n})}\right\| \leq \mathrm{C}_{2},\ \ \ \ \ \ n=1,2,  \notag
\end{align}
and corresponding, respectively, to the outcomes in $\Lambda _{1}$ (on the
side of Alice) and to the outcomes in $\Lambda _{2}$ (on the side of Bob).
\end{itemize}

Suppose that a bipartite quantum system $\mathcal{S}_{q}^{(1)}\mathcal{+S}%
_{q}^{(2)}$ is initially in a separable state and, for concreteness, denote
a separable state by $\rho _{s}.$ Let 
\begin{equation}
\rho _{s}=\sum_{m}\xi _{m}\rho _{1}^{(m)}\otimes \rho _{2}^{(m)},\text{ \ \
\ \ \ \ }\xi _{m}>0,\text{\ \ \ \ \ \ \ }\sum_{m}\xi _{m}=1,  \label{rep}
\end{equation}
be a possible separable representation of $\rho _{s}.$

To any separable representation (\ref{rep}) of $\rho _{s}$, we derive, using
(\ref{sata.aver}) and (\ref{10'}), the following upper bound: 
\begin{align}
\left| \langle \lambda _{1}\lambda _{2}\rangle _{_{\rho _{s}}}^{(\alpha
,\beta _{1})}-\langle \lambda _{1}\lambda _{2}\rangle _{\rho _{s}}^{(\alpha
,\beta _{2})}\right| & \leq \sum_{m}\xi _{m}\mathrm{C}_{1}\left| \mathrm{tr}%
[\rho _{2}^{(m)}W_{2}^{(\beta _{1})}]-\mathrm{tr}[\rho
_{2}^{(m)}W_{2}^{(\beta _{2})}]\right| \text{ }  \label{inequlaity2} \\
& \leq \mathrm{C}_{1}\mathrm{C}_{2}-\frac{\mathrm{C}_{1}}{\mathrm{C}_{2}}%
\sum_{m}\xi _{m}\mathrm{tr}[\rho _{2}^{(m)}W_{2}^{(\beta _{1})}]\mathrm{tr}%
[\rho _{2}^{(m)}W_{2}^{(\beta _{2})}]  \notag \\
& =\mathrm{C}_{1}\mathrm{C}_{2}-\frac{\mathrm{C}_{1}}{\mathrm{C}_{2}}\langle
\lambda _{2}\lambda _{2}^{\prime }\rangle _{\sigma _{2}}^{(\beta _{1},\beta
_{2})},
\end{align}
where:

\noindent \textbf{(i)} $\sigma _{2}$ is the separable density operator 
\begin{equation}
\sigma _{2}:=\sum_{m}\xi _{m}\rho _{2}^{(m)}\otimes \rho _{2}^{(m)}
\label{den}
\end{equation}
on $\mathcal{H}_{2}\otimes \mathcal{H}_{2},$ corresponding to a separable
representation (\ref{rep}) of $\rho _{s}$;

\noindent \textbf{(ii)} $W_{2}^{(\beta _{n})},$ $n=1,2,$ are the
self-adjoint bounded linear operators on $\mathcal{H}_{2},$ defined by (\ref
{operat}), and the notation $\{\cdot \}_{sym}$ is introduced by (\ref{def});

\noindent \textbf{(iii)} $\langle \lambda _{2}\lambda _{2}^{\prime }\rangle
_{\sigma _{2}}^{(\beta _{1},\beta _{2})}$ is the expectation value 
\begin{align}
\langle \lambda _{2}\lambda _{2}^{\prime }\rangle _{\sigma _{2}}^{(\beta
_{1},\beta _{2})}& =\mathrm{tr}[\sigma _{2}\{W_{2}^{(\beta _{1})}\otimes
W_{2}^{(\beta _{2})}\}_{sym}] \\
& =\int_{\mathbb{\Lambda }_{2}\text{ }\times \Lambda _{2}}\lambda
_{2}\lambda _{2}^{\prime }\mathrm{tr}[\sigma _{2}\{M_{2}^{(\beta
_{1})}(d\lambda _{2})\otimes M_{2}^{(\beta _{2})}(d\lambda _{2}^{\prime
})\}_{sym}],  \notag
\end{align}
under the joint experiment, with outcomes in $\Lambda _{2}\times \Lambda
_{2},$ represented by the POV measure 
\begin{equation}
\{M_{2}^{(\beta _{1})}(B_{2})\otimes M_{2}^{(\beta _{2})}(B_{2}^{\prime
})\}_{sym},
\end{equation}
for any subsets $B_{2},B_{2}^{\prime }$ of $\Lambda _{2},$ and performed on
the bipartite quantum system $\mathcal{S}_{q}^{(2)}\mathcal{+S}_{q}^{(2)}$
in the state $\sigma _{2}$.\medskip

\emph{The inequality }(\ref{inequlaity2})\emph{\ establishes the relation
between three\ joint experiments and is valid for any initial separable
state.}

\emph{In general,} \emph{the mean values in the left and the right hand
sides of }(\ref{inequlaity2})\emph{\ refer to joint experiments on different
bipartite systems, namely, on} $\mathcal{S}_{q}^{(1)}\mathcal{+S}_{q}^{(2)}$ 
\emph{and }$\mathcal{S}_{q}^{(2)}\mathcal{+S}_{q}^{(2)},$\emph{\ respectively%
}.\medskip

Moreover, the upper bound in (\ref{inequlaity2}) depends on a chosen
separable representation of $\rho _{s}.$

All possible separable representations of $\rho _{s}$ induce, due to (\ref
{den}), the set 
\begin{equation}
\mathcal{D}_{_{\mathcal{H}_{2}\otimes \mathcal{H}_{2}}}^{(\rho _{s})}
\end{equation}
of density operators $\sigma _{2}$ on $\mathcal{H}_{2}\otimes \mathcal{H}%
_{2}.$ For any two density operators $\sigma _{2}^{^{(1)}}$ and $\sigma
_{2}^{^{(2)}}$ in this set and any non-negative real number $\alpha \leq 1,$
there exists a separable representation of $\rho _{s}$ such the density
operator (\ref{den}), corresponding to this separable representation,
coincides with the convex linear combination 
\begin{equation}
\alpha \sigma _{2}^{^{(1)}}+(1-\alpha )\sigma _{2}^{^{(2)}}.
\end{equation}
Hence, the set $\mathcal{D}_{_{\mathcal{H}_{2}\otimes \mathcal{H}%
_{2}}}^{(\rho _{s})}$ is convex linear.

Furthermore, since the upper bound (\ref{inequlaity2}) is valid for any
separable representation of $\rho _{s},$ we have the following inequality: 
\begin{equation}
\left| \langle \lambda _{1}\lambda _{2}\rangle _{_{\rho _{s}}}^{(\alpha
,\beta _{1})}-\langle \lambda _{1}\lambda _{2}\rangle _{\rho _{s}}^{(\alpha
,\beta _{2})}\right| \leq \inf_{\sigma _{2}\in \mathcal{D}_{_{\mathcal{H}%
_{2}\otimes \mathcal{H}_{2}}}^{(\rho _{s})}}\left\{ \mathrm{C}_{1}\mathrm{C}%
_{2}-\frac{\mathrm{C}_{1}}{\mathrm{C}_{2}}\langle \lambda _{2}\lambda
_{2}^{\prime }\rangle _{\sigma _{2}}^{(\beta _{1},\beta _{2})}\right\} .
\label{inequlaity2'''}
\end{equation}
\bigskip

Consider now the case where both quantum sub-systems $\mathcal{S}_{q}^{(1)}$
and $\mathcal{S}_{q}^{(2)}$ (possibly, different) are described by the same
Hilbert space: $\mathcal{H}_{1}=\mathcal{H}_{2}=\mathcal{H}$.

In this case, if a state $\rho _{s}$ admits a separable representation of
the form 
\begin{equation}
\rho _{s}=\sum_{m}\xi _{m}\rho ^{(m)}\otimes \rho ^{(m)},\text{ \ \ }\xi
_{m}>0,\text{ \ \ }\sum_{m}\xi _{m}=1,  \label{iop}
\end{equation}
then, for this separable representation of $\rho _{s},$ the corresponding
density operator (\ref{den}) coincides with $\rho _{s}$ and the
corresponding upper bound (\ref{inequlaity2}) has the form: 
\begin{equation}
\left| \langle \lambda _{1}\lambda _{2}\rangle _{_{\rho _{s}}}^{(\alpha
,\beta _{1})}-\langle \lambda _{1}\lambda _{2}\rangle _{\rho _{s}}^{(\alpha
,\beta _{2})}\right| \leq \mathrm{C}_{1}\mathrm{C}_{2}-\frac{\mathrm{C}_{1}}{%
\mathrm{C}_{2}}\langle \lambda _{2}\lambda _{2}^{\prime }\rangle _{\rho
_{s}}^{(\beta _{1},\beta _{2})}  \label{xx}
\end{equation}
and refers to the mean values in the same quantum state although to the
joint experiments on different bipartite systems. Namely, in the left-hand
side of (\ref{xx}) the averages correspond to the joint experiments on $%
\mathcal{S}_{q}^{(1)}+\mathcal{S}_{q}^{(2)}$ while in the right hand side on 
$\mathcal{S}_{q}^{(2)}+\mathcal{S}_{q}^{(2)}.\medskip $

Using (\ref{10'}), we further generalize (\ref{inequlaity2}) to the case of
any linear combination of the expectation values: 
\begin{eqnarray}
&&\left| \gamma _{1}\langle \lambda _{1}\lambda _{2}\rangle _{_{\rho
_{s}}}^{(\alpha ,\beta _{1})}+\gamma _{2}\langle \lambda _{1}\lambda
_{2}\rangle _{\rho _{s}}^{(\alpha ,\beta _{2})}\right|  \label{inequlaity2''}
\\
&=&\gamma _{0}\left| \frac{\gamma _{1}}{\gamma _{0}}\langle \lambda
_{1}\lambda _{2}\rangle _{_{\rho _{s}}}^{(\alpha ,\beta _{1})}+\frac{\gamma
_{2}}{\gamma _{0}}\langle \lambda _{1}\lambda _{2}\rangle _{\rho
_{s}}^{(\alpha ,\beta _{2})}\right|  \notag \\
&\leq &\gamma _{0}\mathrm{C}_{1}\sum_{m}\xi _{m}\left| \frac{\gamma _{1}}{%
\gamma _{0}}\mathrm{tr}[\rho _{2}^{(m)}W_{2}^{(\beta _{1})}]+\frac{\gamma
_{2}}{\gamma _{0}}\mathrm{tr}[\rho _{2}^{(m)}W_{2}^{(\beta _{2})}]\right| 
\notag \\
&\leq &\gamma _{0}\mathrm{C}_{1}\mathrm{C}_{2}\left\{ 1+\frac{\gamma
_{1}\gamma _{2}}{\gamma _{0}^{2}\mathrm{C}_{2}^{2}}\text{ }\langle \lambda
_{2}\lambda _{2}^{\prime }\rangle _{\sigma _{2}}^{(\beta _{1},\beta
_{2})}\right\}  \notag \\
&=&\gamma _{0}\mathrm{C}_{1}\mathrm{C}_{2}+\frac{\gamma _{1}\gamma _{2}}{%
\gamma _{0}}\frac{\mathrm{C}_{1}}{\mathrm{C}_{2}}\langle \lambda _{2}\lambda
_{2}^{\prime }\rangle _{\sigma _{2}}^{(\beta _{1},\beta _{2})},  \notag
\end{eqnarray}
where $\gamma _{1},\gamma _{2}$ are any real numbers with $\left| \gamma
_{1}\right| +\left| \gamma _{2}\right| \neq 0$ and $\gamma
_{0}:=\max_{i=1,2}|\gamma _{i}|.$

\subsection{Identical quantum sub-systems}

Consider now the situation where quantum sub-systems are identical: $%
\mathcal{S}_{q}^{(1)}\mathcal{=S}_{q}^{(2)}=\mathcal{S}_{q}$.

In this case, $\mathcal{H}_{1}=\mathcal{H}_{2}=\mathcal{H}$ and, for both
statistics, Boson and Fermi, an initial state must satisfy the relation 
\begin{equation}
\mathrm{S}_{2}\rho =\rho ,  \label{sym}
\end{equation}
where we denote by $\mathrm{S}_{2}$ the symmetrization operator on $\mathcal{%
H}\otimes \mathcal{H}$ (see [8], page 53).

Moreover, for any joint experiment on $\mathcal{S}_{q}+\mathcal{S}_{q}$ of
the Alice/Bob type$,$ with outcomes in $\Lambda _{1}\times \Lambda _{2}$,
the POV measure of each individual (marginal) experiment on the side of
Alice or Bob must have a symmetrized tensor product form and be specified by
a set $\Lambda _{i}$, $i=1,2,$ of outcomes but not by the ''side'' of the
tensor product.

The latter means that, for this type of a joint experiment on $\mathcal{S}%
_{q}+\mathcal{S}_{q},$ the POV measure must have the following form: 
\begin{equation}
M(B_{1}\times B_{2})=\{M_{1}(B_{1})\otimes M_{2}(B_{2})\}_{sym}=\frac{1}{2}%
\{M_{1}(B_{1})\otimes M_{2}(B_{2})+M_{2}(B_{2})\otimes M_{1}(B_{1})\},
\label{se}
\end{equation}
for any subset $B_{1}$ of $\Lambda _{1}$ and any subset $B_{2}$ of $\Lambda
_{2}.$

However, for further calculation of traces in a state $\rho ,$ satisfying
the condition (\ref{sym}), the symmetrization (\ref{se}) is not essential
since, for this state, 
\begin{equation}
\mathrm{tr}[\rho \{V_{1}\otimes V_{2}\}_{sym}]=\mathrm{tr}[\rho
(V_{1}\otimes V_{2})],
\end{equation}
for any bounded $V_{1}$ and $V_{2}$ on $\mathcal{H}.$

Consider two joint experiments (of the Alice/Bob type) on $\mathcal{S}_{q}+%
\mathcal{S}_{q}$, represented by the POV measures 
\begin{align}
M^{(\alpha ,\beta _{1})}(B_{1}\times B_{2})& =\{M_{1}^{(\alpha
)}(B_{1})\otimes M_{2}^{(\beta _{1})}(B_{2})\}_{sym},  \label{13'} \\
M^{(\alpha ,\beta _{2})}(B_{1}\times B_{2})& =\{M_{1}^{(\alpha
)}(B_{1})\otimes M_{2}^{(\beta _{2})}(B_{2})\}_{sym},  \notag
\end{align}
and performed on a bipartite quantum system $\mathcal{S}_{q}\mathcal{+S}_{q}$
being initially in a state $\rho ,$ satisfying (\ref{sym}).

In (\ref{13'}), the parameters $\alpha $ and $\beta $ (of any nature)
characterize set-ups of the corresponding (marginal) experiments with
outcomes in $\Lambda _{1}$ and in $\Lambda _{2}$, respectively.

For each of the joint experiments (\ref{13'}), the formula 
\begin{align}
\langle \lambda _{1}\lambda _{2}\rangle _{_{\rho }}^{(\alpha ,\beta _{n})}&
=\int \lambda _{1}\lambda _{2}\mathrm{tr}[\rho \{M_{1}^{(\alpha )}(d\lambda
_{1})\otimes M_{2}^{(\beta _{n})}(d\lambda _{2})\}_{sym}]  \label{ji} \\
& =\mathrm{tr}[\rho \{W_{1}^{(\alpha )}\otimes W_{2}^{(\beta
_{n})}\}_{sym}],\ \ \ \ n=1,2,  \notag
\end{align}
represents the expectation value of the product $\lambda _{1}\lambda _{2}$
of the observed outcomes. Here, $W_{1}^{(\alpha )}$ and $W_{2}^{(\beta
_{n})} $ are the self-adjoint bounded linear operators on $\mathcal{H},$
defined by (\ref{operat}) and corresponding to the observed outcomes in $%
\Lambda _{1}$ (on the side of Alice) and in $\Lambda _{2},($on the side of
Bob), respectively.\medskip

Suppose that a bipartite quantum system $\mathcal{S}_{q}\mathcal{+S}_{q}$ is
initially in\ a separable quantum state $\rho _{s}$ and let 
\begin{equation}
\rho _{s}=\sum_{m}\frac{\xi _{m}}{2}\{\rho _{m}\otimes \rho _{m}^{\prime
}+\rho _{m}^{\prime }\otimes \rho _{m}\}=\sum_{m}\xi _{m}\{\rho _{m}\otimes
\rho _{m}^{\prime }\}_{sym},\text{ \ \ \ }\xi _{m}>0,\text{\ \ \ \ }%
\sum_{m}\xi _{m}=1,  \label{hg}
\end{equation}
be a separable representation of $\rho _{s}.$

To any separable representation (\ref{hg}) of $\rho _{s},$ we derive,
similarly to (\ref{inequlaity2}), the following quantum inequality: 
\begin{eqnarray}
&&\left| \langle \lambda _{1}\lambda _{2}\rangle _{\rho _{s}}^{(\alpha
,\beta _{1})}-\langle \lambda _{1}\lambda _{2}\rangle _{\rho _{s}}^{(\alpha
,\beta _{2})}\right|  \label{inequlaity2'} \\
&\leq &\sum_{m}\frac{\xi _{m}}{2}\mathrm{C}_{1}\left\{ \left| \mathrm{tr}%
[\rho _{m}W_{2}^{(\beta _{1})}]-\mathrm{tr}[\rho _{m}W_{2}^{(\beta
_{2})}]\right| +\left| \mathrm{tr}[\rho _{m}^{\prime }W_{2}^{(\beta _{1})}]-%
\mathrm{tr}[\rho _{m}^{\prime }W_{2}^{(\beta _{2})}]\right| \right\}  \notag
\\
&\leq &\mathrm{C}_{1}\mathrm{C}_{2}-\frac{\mathrm{C}_{1}}{\mathrm{C}_{2}}%
\sum_{m}\frac{\xi _{m}}{2}\left\{ \mathrm{tr}[\rho _{m}W_{2}^{(\beta _{1})}]%
\text{ }\mathrm{tr}[\rho _{m}W_{2}^{(\beta _{2})}]+\mathrm{tr}[\rho
_{m}^{\prime }W_{2}^{(\beta _{1})}]\text{ }\mathrm{tr}[\rho _{m}^{\prime
}W_{2}^{(\beta _{2})}]\right\}  \notag \\
&\leq &\mathrm{C}_{1}\mathrm{C}_{2}-\frac{\mathrm{C}_{1}}{\mathrm{C}_{2}}%
\langle \lambda _{2}\lambda _{2}^{\prime }\rangle _{\sigma }^{(\beta
_{1},\beta _{2})},  \notag
\end{eqnarray}
where 
\begin{equation}
\sigma =\sum_{m}\frac{\xi _{m}}{2}\{\rho _{m}\otimes \rho _{m}+\rho
_{m}^{\prime }\otimes \rho _{m}^{\prime }\}  \label{56}
\end{equation}
is the separable density operator on $\mathcal{H}\otimes \mathcal{H},$
corresponding to a separable representation (\ref{hg}) of $\rho _{s},$ and 
\begin{eqnarray}
\langle \lambda _{2}\lambda _{2}^{\prime }\rangle _{\sigma }^{(\beta
_{1},\beta _{2})} &=&\mathrm{tr}[\sigma \{W_{2}^{(\beta _{1})}\otimes
W_{2}^{(\beta _{2})}\}_{sym}]  \label{3} \\
&=&\int \lambda _{2}\lambda _{2}^{\prime }\mathrm{tr}[\sigma \{M_{2}^{(\beta
_{1})}(d\lambda _{2})\otimes M_{2}^{(\beta _{2})}(d\lambda _{2}^{\prime
})\}_{sym}]  \notag
\end{eqnarray}
is the expectation value under the joint experiment, with outcomes in $%
\Lambda _{2}\times \Lambda _{2}$ (both outcomes are on the side of Bob),
described by the POV measure 
\begin{equation}
\{M_{2}^{(\beta _{1})}(B_{2})\otimes M_{2}^{(\beta _{2})}(B_{2}^{\prime
})\}_{sym},
\end{equation}
for any subsets $B_{2},B_{2}^{\prime }$ of $\Lambda _{2},$ and performed on $%
\mathcal{S}_{q}\mathcal{+S}_{q}$ in the state $\sigma ,$ possibly different
from $\rho _{s}.$

Notice that, in the state $\sigma ,$ the correlation function $\langle
\lambda _{2}\lambda _{2}^{\prime }\rangle _{\sigma }^{(\beta _{1},\beta
_{1})}$ is always positive: 
\begin{equation}
\langle \lambda _{2}\lambda _{2}^{\prime }\rangle _{\sigma }^{(\beta
_{1},\beta _{1})}\geq 0.  \label{sdf}
\end{equation}

Quite similarly to our derivation of (\ref{inequlaity2''}), for any
separable state $\rho _{s},$ we have: 
\begin{equation}
\left| \gamma _{1}\langle \lambda _{1}\lambda _{2}\rangle _{_{\rho
_{s}}}^{(\alpha ,\beta _{1})}+\gamma _{2}\langle \lambda _{1}\lambda
_{2}\rangle _{\rho _{s}}^{(\alpha ,\beta _{2})}\right| \leq \gamma _{0}%
\mathrm{C}_{1}\mathrm{C}_{2}+\frac{\gamma _{1}\gamma _{2}}{\gamma _{0}}\frac{%
\mathrm{C}_{1}}{\mathrm{C}_{2}}\langle \lambda _{2}\lambda _{2}^{\prime
}\rangle _{\sigma }^{(\beta _{1},\beta _{2})},
\end{equation}
where $\gamma _{1},\gamma _{2}$ are any real numbers with $\left| \gamma
_{1}\right| +\left| \gamma _{2}\right| \neq 0$ and $\gamma
_{0}:=\max_{i=1,2}|\gamma _{i}|.$

\subsubsection{Quantum analogues of Bell's inequality}

Consider further the case where, in (\ref{hu}), $\mathrm{C}_{1}=\mathrm{C}%
_{2}=\mathrm{C.}$

\emph{Assume} that the (marginal) experiments on the sides of Alice and Bob
are similar - in the sense that 
\begin{eqnarray}
\int \lambda _{2}M_{2}^{(\beta _{1})}(d\lambda _{2}) &=&\int \lambda
_{1}M_{1}^{(\beta _{1})}(d\lambda _{1})\text{ \ }  \label{2} \\
\text{\ \ } &\Leftrightarrow &\text{ \ \ }W_{2}^{(\beta _{1})}=W_{1}^{(\beta
_{1})}.  \notag
\end{eqnarray}
\emph{The condition }(\ref{2})\emph{\ does not represent the Bell
correlation restrictions }[2-4]\emph{\ on the observed outcomes on the sides
of Alice and Bob and is usually fulfilled}\footnote{%
For identical quantum sub-systems, this is a condition on identical
measurement devices, used on both sides, for example, identical polarization
analyzers.}\emph{\ under Alice/Bob joint quantum experiments.}

Under the condition (\ref{2}), the expectation value 
\begin{eqnarray}
\langle \lambda _{2}\lambda _{2}^{\prime }\rangle _{\sigma }^{(\beta
_{1},\beta _{2})} &=&\mathrm{tr}[\sigma \{W_{2}^{(\beta _{1})}\otimes
W_{2}^{(\beta _{2})}\}_{sym}] \\
&=&\mathrm{tr}[\sigma \left\{ \int \lambda _{2}\lambda _{2}^{\prime
}\{M_{2}^{(\beta _{1})}(d\lambda _{2})\otimes M_{2}^{(\beta _{2})}(d\lambda
_{2}^{\prime })\}_{sym}\right\} ]  \notag \\
&=&\int \lambda _{1}\lambda _{2}^{\prime }\mathrm{tr}[\sigma \{M_{1}^{(\beta
_{1})}(d\lambda _{1})\otimes M_{2}^{(\beta _{2})}(d\lambda _{2}^{\prime
})\}_{sym}]  \notag \\
&=&\langle \lambda _{1}\lambda _{2}\rangle _{\sigma }^{(\beta _{1},\beta
_{2})},  \notag
\end{eqnarray}
and, hence, the inequality (\ref{inequlaity2'}) takes the Bell-form: 
\begin{equation}
\left| \langle \lambda _{1}\lambda _{2}\rangle _{\rho _{s}}^{(\alpha ,\beta
_{1})}-\langle \lambda _{1}\lambda _{2}\rangle _{\rho _{s}}^{(\alpha ,\beta
_{2})}\right| \leq \mathrm{C}^{2}-\langle \lambda _{1}\lambda _{2}\rangle
_{\sigma }^{(\beta _{1},\beta _{2})},  \label{new}
\end{equation}
with \emph{all three mean values referring to Alice/Bob joint experiments.}

However, in general, $\sigma \neq \rho _{s}\ $and this means that, under
these joint experiments, initial states of $\mathcal{S}_{q}+\mathcal{S}_{q}$
may be different\emph{.}

We call (\ref{new})\emph{\ a quantum analogue of the original Bell
inequality for a separable quantum state.\medskip }

To different separable representations of a separable state $\rho _{s},$
there correspond different terms $\langle \lambda _{1}\lambda _{2}\rangle
_{\sigma }^{(\beta _{1},\beta _{2})}$ in the right-hand side of (\ref{new}).

In general, for a separable state, any quantum inequality (\ref{new}) may
not coincide with the original Bell inequality and, hence, for this state,
the original Bell inequality may be violated.\medskip

\emph{Let specify the conditions under which a separable state of }$\mathcal{%
S}_{q}+\mathcal{S}_{q}$ \emph{satisfies the original Bell inequality. }We
suppose that the condition (\ref{2}) is fulfilled.\bigskip

\begin{enumerate}
\item  Let an initial separable state $\rho _{s}$ on $\mathcal{H}\otimes 
\mathcal{H}$ admit a representation of the special form 
\begin{equation}
\rho _{s}=\sum_{m}\xi _{m}\rho _{m}\otimes \rho _{m},\text{ \ \ \ }\xi
_{m}>0,\text{\ \ \ }\sum_{m}\xi _{m}=1.  \label{jo}
\end{equation}
Then, from (\ref{56}) it follows that, for this separable representation, 
\begin{equation}
\sigma =\rho _{s}
\end{equation}
and, hence, the corresponding quantum inequality (\ref{new}) reduces to 
\begin{equation}
\left| \langle \lambda _{1}\lambda _{2}\rangle _{\rho _{s}}^{(\alpha ,\beta
_{1})}-\langle \lambda _{1}\lambda _{2}\rangle _{\rho _{s}}^{(\alpha ,\beta
_{2})}\right| \leq \mathrm{C}^{2}-\langle \lambda _{1}\lambda _{2}\rangle
_{\rho _{s}}^{(\beta _{1},\beta _{2})},  \label{lub}
\end{equation}

so that \emph{a separable state }(\ref{jo})\emph{\ satisfies the original
Bell inequality in its perfect correlation form.}

For a separable state (\ref{jo}), the correlation function $\langle \lambda
_{1}\lambda _{2}\rangle _{\rho _{s}}^{(\beta _{1},\beta _{1})}$ is always
non-negative: 
\begin{equation}
\langle \lambda _{1}\lambda _{2}\rangle _{\rho _{s}}^{(\beta _{1},\beta
_{1})}\geq 0
\end{equation}
and may take any value in $[0,\mathrm{C}^{2}].\medskip $

\item  Consider further the situation where the marginal experiments and a
separable state $\rho _{s}$ are such that for some separable representation (%
\ref{hg}) of $\rho _{s}:$ 
\begin{equation}
\mathrm{tr}[\sigma \{W_{1}^{(\beta _{1})}\otimes W_{2}^{(\beta
_{2})}\}_{sym}]=\pm \mathrm{tr}[\rho _{s}\{W_{1}^{(\beta _{1})}\otimes
W_{2}^{(\beta _{2})}\}_{sym}],  \label{sir}
\end{equation}
or, equivalently, 
\begin{equation}
\langle \lambda _{1}\lambda _{2}\rangle _{\sigma }^{(\beta _{1},\beta
_{2})}=\pm \langle \lambda _{1}\lambda _{2}\rangle _{\rho _{s}}^{(\beta
_{1},\beta _{2})},  \label{vbi}
\end{equation}
Notice, that, under the condition (\ref{jo}) in the point 1, the condition (%
\ref{sir}) is satisfied. However, we have specially separated these two
cases since (\ref{jo}) represents a restriction only on a separable state
while (\ref{sir}) is, in general, a restriction on the combination - a joint
experiment\ plus a state.

Since, in (\ref{vbi}), in the state $\sigma $ the correlation function $%
\langle \lambda _{1}\lambda _{2}\rangle _{\sigma }^{(\beta _{1},\beta _{1})}$
is always non-negative (see (\ref{sdf}), the necessary condition for (\ref
{vbi}) to hold constitutes 
\begin{equation}
\pm \langle \lambda _{1}\lambda _{2}\rangle _{\rho _{s}}^{(\beta _{1},\beta
_{1})}\geq 0.
\end{equation}
Hence, the signs ''plus'' or ''minus'' in (\ref{vbi}) coincide with the
corresponding signs of the correlation function 
\begin{equation}
\langle \lambda _{1}\lambda _{2}\rangle _{\rho _{s}}^{(\beta _{1},\beta
_{1})}.
\end{equation}

Under the condition (\ref{vbi}), the corresponding quantum analogue (\ref
{new}) reduces to 
\begin{equation}
\left| \langle \lambda _{1}\lambda _{2}\rangle _{\rho _{s}}^{(\alpha ,\beta
_{1})}-\langle \lambda _{1}\lambda _{2}\rangle _{\rho _{s}}^{(\alpha ,\beta
_{2})}\right| \leq \mathrm{C}^{2}\mp \text{ }\langle \lambda _{1}\lambda
_{2}\rangle _{\rho _{s}}^{(\beta _{1},\beta _{2})},
\end{equation}
and \emph{coincides with the original Bell inequality, in its perfect
correlation or anti-correlation forms}.

In particular, (\ref{vbi}) (equivalently, (\ref{sir})) is satisfied if, for
all indices $m,$ 
\begin{equation}
\mathrm{tr}[\rho _{m}W_{1}^{(\beta _{1})}]=\pm \mathrm{tr}[\rho _{m}^{\prime
}W_{1}^{(\beta _{1})}].  \label{sor}
\end{equation}
For a separable state, \emph{Bell's correlation restrictions} 
\begin{equation}
\langle \lambda _{1}\lambda _{2}\rangle _{\rho _{s}}^{(\beta _{1},\beta
_{1})}=\pm 1,
\end{equation}
introduced in [4], sections 2, 4, for the derivation of the original Bell
inequality in the frame of a LHV model, \emph{represent a particular case of}
the condition (\ref{sor}), and, hence, of the condition (\ref{vbi}).\bigskip
\end{enumerate}

\begin{example}
Consider the example of section 2. In this example, 
\begin{equation}
\mathrm{C}=1,\text{ \ \ }\rho _{s}=\rho _{0},\text{ \ \ }\alpha =\theta _{a},%
\text{ \ }\beta _{1}=\theta _{b},\text{ \ \ }\beta _{2}=\theta _{c}.
\end{equation}
For any parameters $\theta _{a},$ $\theta _{b},$ $\theta _{c},$ we have: 
\begin{align}
\langle \lambda _{1}\lambda _{2}\rangle _{\rho _{0}}^{(\theta _{a},\theta
_{b})}& =\mathrm{tr}[\rho _{0}\{J^{(\theta _{a})}\otimes J^{(\theta
_{b})}\}_{sym}]=-cos2\theta _{a}\text{ }cos2\theta _{b},  \label{ui} \\
\langle \lambda _{1}\lambda _{2}\rangle _{\rho _{0}}^{(\theta _{a},\theta
_{c})}& =\mathrm{tr}[\rho _{0}\{J^{(\theta _{a})}\otimes J^{(\theta
_{c})}\}_{sym}]=-cos2\theta _{a}\text{ }cos2\theta _{c},  \notag \\
\sigma & =\frac{1}{2}\{\left\{ |\uparrow ><\uparrow |\otimes |\uparrow
><\uparrow |+|\downarrow ><\downarrow |\otimes |\downarrow ><\downarrow
|\right\} ,  \notag \\
\langle \lambda _{1}\lambda _{2}\rangle _{\sigma }^{(\theta _{b},\theta
_{c})}& =\mathrm{tr}[\sigma \{J^{(\theta _{b})}\otimes J^{(\theta
_{c})}\}_{sym}]=cos2\theta _{b}\text{ }cos2\theta _{c}.  \notag
\end{align}
For any $\theta _{b},$ $\theta _{c},$%
\begin{equation}
\langle \lambda _{1}\lambda _{2}\rangle _{\sigma }^{(\theta _{b},\theta
_{c})}=-\langle \lambda _{1}\lambda _{2}\rangle _{\rho _{0}}^{(\theta
_{b},\theta _{c})},
\end{equation}
and hence, the condition (\ref{vbi}) is satisfied. For all $\theta _{a},$ $%
\theta _{b},$ $\theta _{c},$ the quantum inequality (\ref{new}) is given by 
\begin{equation}
\left| \langle \lambda _{1}\lambda _{2}\rangle _{\rho _{0}}^{(\theta
_{a},\theta _{b})}-\langle \lambda _{1}\lambda _{2}\rangle _{\rho
_{0}}^{(\theta _{a},\theta _{c})}\right| \leq \mathrm{1}+\langle \lambda
_{1}\lambda _{2}\rangle _{\rho _{0}}^{(\theta _{b},\theta _{c})},
\end{equation}
and coincides with the anti-correlation form of the original Bell inequality.
\end{example}

\bigskip

\emph{Thus, under the condition }(\ref{2})\emph{\ on similarity}\footnote{%
We would like to underline once more that this is not a condition on any
correlation between the observed outcomes on the sides of Alice and Bob.}%
\emph{\ of experimental devices on the side of Alice and the side of Bob,
any separable quantum state of }$\mathcal{S}_{q}+\mathcal{S}_{q}$ \emph{%
satisfies a quantum analogue of the original Bell inequality.}

\emph{Under the sufficient conditions, specified in items 1 and 2, a
separable state satisfies the original Bell inequality, in its perfect
correlation or anti-correlation forms.}

\section{Extended CHSH inequality}

In this section, we introduce the extended CHSH inequality for any linear
combination of mean values. Based on our results in section 3.1, we prove
that this inequality is valid for any separable state.

Consider four joint experiments of the Alice/Bob type on a bipartite quantum
system $\mathcal{S}_{q}^{(1)}\mathcal{+S}_{q}^{(2)}$ on $\mathcal{H}%
_{1}\otimes \mathcal{H}_{2}.$ Let all these experiments have outcomes in $%
\Lambda _{1}\times \Lambda _{2}$ (see (\ref{hu})) and be described by the
POV measures: 
\begin{align}
M^{(a,b)}(B_{1}\times B_{2})& =M_{1}^{(a)}(B_{1})\otimes M_{2}^{(b)}(B_{2}),%
\text{ \ \ \ \ \ \ }M^{(c,b)}(B_{1}\times B_{2})=M_{1}^{(c)}(B_{1})\otimes
M_{2}^{(b)}(B_{2}),  \label{io} \\
M^{(c,d)}(B_{1}\times B_{2})& =M_{1}^{(c)}(B_{1})\otimes M_{2}^{(d)}(B_{2}),%
\text{ \ \ \ \ \ \ }M^{(a,d)}(B_{1}\times B_{2})=M_{1}^{(a)}(B_{1})\otimes
M_{2}^{(d)}(B_{2}),  \notag
\end{align}
for any subset $B_{1}\subseteq \Lambda _{1}$ and any subset $B_{2}\subseteq
\Lambda _{2}.$

In (\ref{io}), the parameters $a,b,$ $c,$ $d$ are of any nature and $a,b$
refer to the set-ups of the experiments with outcomes in $\Lambda _{1}$ (the
''side'' of Alice) while $b,d$ refer to the set-ups of the experiments with
outcomes in $\Lambda _{2}$ (the ''side'' of Bob).

Suppose that all four joint experiments (\ref{io}) are performed on a
bipartite quantum system $\mathcal{S}_{q}^{(1)}\mathcal{+S}_{q}^{(2)}$ in
the same separable state $\rho _{s}.$

For any real numbers $\gamma _{i},$ $i=1,...,4,$ with 
\begin{equation}
\left| \gamma _{1}\right| +\left| \gamma _{2}\right| +\left| \gamma
_{3}\right| +\left| \gamma _{4}\right| \neq 0,
\end{equation}
let estimate the linear combination 
\begin{equation}
\left| \gamma _{1}\langle \lambda _{1}\lambda _{2}\rangle _{\rho
_{s}}^{(a,b)}+\gamma _{2}\langle \lambda _{1}\lambda _{2}\rangle _{\rho
_{s}}^{(c,b)}+\gamma _{3}\langle \lambda _{1}\lambda _{2}\rangle _{\rho
_{s}}^{(c,d)}+\gamma _{4}\langle \lambda _{1}\lambda _{2}\rangle _{\rho
_{s}}^{(a,d)}\right|
\end{equation}
of the mean values under four joint experiments (\ref{io}).

Similarly to the derivation of (\ref{inequlaity2''}), to any separable
representation (\ref{rep}) of $\rho _{s},$ we have: 
\begin{align}
& \left| \gamma _{1}\langle \lambda _{1}\lambda _{2}\rangle _{\rho
_{s}}^{(a,b)}+\gamma _{2}\langle \lambda _{1}\lambda _{2}\rangle _{\rho
_{s}}^{(c,b)}+\gamma _{3}\langle \lambda _{1}\lambda _{2}\rangle _{\rho
_{s}}^{(c,d)}+\gamma _{4}\langle \lambda _{1}\lambda _{2}\rangle _{\rho
_{s}}^{(a,d)}\right|  \label{cs} \\
& \leq \left| \gamma _{1}\langle \lambda _{1}\lambda _{2}\rangle _{\rho
_{s}}^{(a,b)}+\gamma _{4}\langle \lambda _{1}\lambda _{2}\rangle _{\rho
_{s}}^{(a,d)}\right| +\left| \gamma _{2}\langle \lambda _{1}\lambda
_{2}\rangle _{\rho _{s}}^{(c,b)}+\gamma _{3}\langle \lambda _{1}\lambda
_{2}\rangle _{\rho _{s}}^{(c,d)}\right|  \notag \\
& =\widetilde{\gamma }_{0}\left| \frac{\gamma _{1}}{\widetilde{\gamma }_{0}}%
\langle \lambda _{1}\lambda _{2}\rangle _{\rho _{s}}^{(a,b)}+\frac{\gamma
_{4}}{\widetilde{\gamma }_{0}}\langle \lambda _{1}\lambda _{2}\rangle _{\rho
_{s}}^{(a,d)}\right| +\widetilde{\gamma }_{0}\left| \frac{\gamma _{2}}{%
\widetilde{\gamma }_{0}}\langle \lambda _{1}\lambda _{2}\rangle _{\rho
_{s}}^{(c,b)}+\frac{\gamma _{3}}{\widetilde{\gamma }_{0}}\langle \lambda
_{1}\lambda _{2}\rangle _{\rho _{s}}^{(c,d)}\right|  \notag \\
& \leq 2\widetilde{\gamma }_{0}\mathrm{C}_{1}\mathrm{C}_{2}+\frac{C_{1}}{%
\widetilde{\gamma }_{0}C_{2}}\left\{ \gamma _{1}\gamma _{4}+\gamma
_{2}\gamma _{3}\right\} \langle \lambda _{2}\lambda _{2}^{\prime }\rangle
_{\sigma _{2}}^{(b,d)},  \notag
\end{align}
where $\widetilde{\gamma }_{0}:=\max_{i=1,...,4}|\gamma _{i}|$ and 
\begin{equation}
\langle \lambda _{2}\lambda _{2}^{\prime }\rangle _{\sigma _{2}}^{(b,d)}:=%
\mathrm{tr}[\sigma _{2}\{W_{2}^{(b)}\otimes W_{2}^{(d)}\}_{sym}],  \notag
\end{equation}
with the density operator $\sigma _{2}$ on $\mathcal{H}_{2}\otimes \mathcal{H%
}_{2}$, defined by (\ref{den}).

However, if, in the second line of (\ref{cs}), we combine the terms in
another way, we derive: 
\begin{eqnarray}
&&\left| \text{ }\gamma _{1}\langle \lambda _{1}\lambda _{2}\rangle _{\rho
_{s}}^{(a,b)}+\gamma _{2}\langle \lambda _{1}\lambda _{2}\rangle _{\rho
_{s}}^{(c,b)}+\gamma _{3}\langle \lambda _{1}\lambda _{2}\rangle _{\rho
_{s}}^{(c,d)}+\gamma _{4}\langle \lambda _{1}\lambda _{2}\rangle _{\rho
_{s}}^{(a,d)}\text{ }\right|  \label{cs'} \\
&\leq &\left| \text{ }\gamma _{1}\langle \lambda _{1}\lambda _{2}\rangle
_{\rho _{s}}^{(a,b)}+\gamma _{2}\langle \lambda _{1}\lambda _{2}\rangle
_{\rho _{s}}^{(c,b)}\text{ }\right| +\left| \text{ }\gamma _{3}\langle
\lambda _{1}\lambda _{2}\rangle _{\rho _{s}}^{(c,d)}+\gamma _{4}\langle
\lambda _{1}\lambda _{2}\rangle _{\rho _{s}}^{(a,d)}\text{ }\right|  \notag
\\
&\leq &2\widetilde{\gamma }_{0}\mathrm{C}_{1}\mathrm{C}_{2}+\frac{C_{1}}{%
\widetilde{\gamma }_{0}C_{2}}\left\{ \gamma _{1}\gamma _{2}+\gamma
_{3}\gamma _{4}\right\} \langle \lambda _{1}\lambda _{1}^{\prime }\rangle
_{\sigma _{1}}^{(a,c)},  \notag
\end{eqnarray}
where 
\begin{equation}
\langle \lambda _{1}\lambda _{1}^{\prime }\rangle _{\sigma _{1}}^{(a,c)}=%
\mathrm{tr}[\sigma _{1}\{W_{1}^{(a)}\otimes W_{1}^{(c)}\}_{sym}],  \notag
\end{equation}
with \ 
\begin{equation}
\sigma _{1}=\sum_{m}\xi _{m}\rho _{1}^{(m)}\otimes \rho _{1}^{(m)}
\end{equation}
being the density operator on $\mathcal{H}_{1}\otimes \mathcal{H}_{1},$
corresponding to a separable representation (\ref{rep}).\medskip

From (\ref{cs}) and (\ref{cs'}) it follows that, for any separable state $%
\rho _{s},$ the inequality 
\begin{equation}
\left| \gamma _{1}\langle \lambda _{1}\lambda _{2}\rangle _{\rho
_{s}}^{(a,b)}+\gamma _{2}\langle \lambda _{1}\lambda _{2}\rangle _{\rho
_{s}}^{(c,b)}+\gamma _{3}\langle \lambda _{1}\lambda _{2}\rangle _{\rho
_{s}}^{(c,d)}+\gamma _{4}\langle \lambda _{1}\lambda _{2}\rangle _{\rho
_{s}}^{(a,d)}\right| \leq 2\widetilde{\gamma }_{0}\mathrm{C}_{1}\mathrm{C}%
_{2}.  \label{cs2}
\end{equation}
holds for all real numbers $\gamma _{1},$ $\gamma _{2},$ $\gamma _{3},$ $%
\gamma _{4}$, with $\left| \gamma _{1}\right| +\left| \gamma _{2}\right|
+\left| \gamma _{3}\right| +\left| \gamma _{4}\right| \neq 0,$ such that 
\begin{equation}
\text{\ }\gamma _{1}\gamma _{4}=-\gamma _{2}\gamma _{3}\text{ \ \ or \ \ }%
\gamma _{1}\gamma _{2}=-\gamma _{3}\gamma _{4}.  \label{28}
\end{equation}
\medskip

\emph{We refer to }(\ref{cs2}) \emph{as the} \emph{extended} \emph{CHSH
inequality}. The original CHSH inequality 
\begin{equation}
\left| \text{ }\langle \lambda _{1}\lambda _{2}\rangle _{\rho
_{s}}^{(a,b)}+\langle \lambda _{1}\lambda _{2}\rangle _{\rho
_{s}}^{(c,b)}+\langle \lambda _{1}\lambda _{2}\rangle _{\rho
_{s}}^{(c,d)}-\langle \lambda _{1}\lambda _{2}\rangle _{\rho _{s}}^{(a,d)}%
\text{ }\right| \leq 2  \label{hi'}
\end{equation}
is the special case of (\ref{cs2}) for 
\begin{equation}
\gamma _{1}=\gamma _{2}=\gamma _{3}=-\gamma _{4},\text{\ \ \ \ \ \ \ }%
\mathrm{C}_{1}\mathrm{C}_{2}=1.
\end{equation}

\emph{Similarly to our presentation in this section, it is easy to verify
that the extended CHSH inequality is valid for all joint classical
measurements}\footnote{%
For the description of a classical measurement, see appendix.}\emph{.}

\emph{Thus, in contrast to the situation with the original Bell inequality }(%
\ref{34})\emph{, such a classical constraint as the extended CHSH inequality
is valid for any separable quantum state. Moreover, this inequality holds
for a variety of linear combinations of the mean values.} \bigskip

\noindent \textbf{Acknowledgments.}{\Large \ }I am indebted to Klaus Molmer,
Asher Peres, Michael Steiner and Marek Zukowski for valuable comments and
useful discussions.\bigskip

\section{Appendix. Classical measurements}

Let a system $\mathcal{S}$ be described in terms of some parameters $\theta
\in \Theta $ (hidden or real) and a probability distribution $\pi $ of these
parameters.

An experiment, constituting a non-perturbing measurement of some property $A$
of $\mathcal{S}$, existed before an experiment, is described by a measurable
function\footnote{%
In classical probability, these functions are called random variables.} $%
f_{A}$ on $\Theta ,$ with values that are outcomes under this experiment.
This type of an experiment is called (see in [9, 10]) a classical
measurement.

Under this type of an \emph{ideal} experiment on $\mathcal{S}$, the
probability distribution of outcomes is an image of an initial probability
distribution $\pi $ and does not depend on an arrangement of a measurement.
Notice that Bell's LHV model ([4], sections 2, 4) describes a perturbing
classical experiment, where the probabilities of the system properties,
existed before an experiment, are modified by this measurement.

Any joint classical measurement on two system properties (say $A$ and $D)$
is described by two real-valued functions $f_{A},$ $f_{D}$ on $\Theta $,
with values equal to the outcomes $\lambda _{1}$ and $\lambda _{2},$
observed under this joint classical measurement.

\emph{Notice also that, for the probabilistic description of any joint
experiment, it is not essential whether or not individual (i.e. marginal)
experiments are separated in time or space.}

The expectation value $\langle \lambda _{1}\lambda _{2}\rangle
_{cl}^{(A\&D)} $ of the product of the observed outcomes is given by 
\begin{equation}
\langle \lambda _{1}\lambda _{2}\rangle _{cl}^{(A\&D)}=\int_{\Theta
}f_{A}(\theta )f_{D}(\theta )\pi (d\theta ).
\end{equation}

Suppose now that we have two joint classical measurements of properties $%
A\&D_{1}$ and $A\&D_{2}$ and 
\begin{equation}
|f_{A}(\theta )|\leq \mathrm{C}_{1},\text{ \ \ \ \ \ }\left|
f_{D_{1}}(\theta )\right| \leq \mathrm{C}_{2},\text{ \ \ \ \ \ }\left|
f_{D_{2}}(\theta )\right| \leq \mathrm{C}_{2}\text{,}
\end{equation}
for all those $\theta \in \Theta $ where $\pi $ does not vanish.

Consider, in a very general setting, the relation between the expectation
values 
\begin{eqnarray}
\langle \lambda _{1}\lambda _{2}\rangle _{cl}^{(A\&D_{1})} &=&\int_{\Theta
}f_{A}(\theta )f_{D_{1}}(\theta )\pi (d\theta ),\text{ \ \ \ \ \ }\langle
\lambda _{1}\lambda _{2}\rangle _{cl}^{(A\&D_{2})}=\int_{\Theta
}f_{A}(\theta )\text{\ }f_{D_{2}}(\theta )\pi (d\theta ), \\
\langle \lambda _{1}\lambda _{2}\rangle _{cl}^{(D_{1}\&D_{2})}
&=&\int_{\Theta }f_{D_{1}}(\theta )f_{D_{2}}(\theta )\pi (d\theta ),  \notag
\end{eqnarray}
under three joint classical measurements on properties 
\begin{equation}
A\&D_{1},\text{ \ \ \ \ }A\&D_{2}\text{, \ \ \ \ }D_{1}\&D_{2}
\end{equation}
of $\mathcal{S}.$ Due to the inequality (\ref{10'}), we have: 
\begin{eqnarray}
&&\left| \langle \lambda _{1}\lambda _{2}\rangle _{cl}^{(A\&D_{1})}-\langle
\lambda _{1}\lambda _{2}\rangle _{cl}^{(A\&D_{2})}\right|  \label{35} \\
&\leq &\int_{\Theta }|f_{A}(\theta )|\text{\ }\left| \text{ }%
f_{D_{1}}(\theta )-f_{D_{2}}(\theta )\right| \pi (d\theta )  \notag \\
&\leq &\mathrm{C}_{1}\int_{\Theta }\left| \text{ }f_{D_{1}}(\theta
)-f_{D_{2}}(\theta )\text{ }\right| \pi (d\theta )  \notag \\
&\leq &\mathrm{C}_{1}\mathrm{C}_{2}-\frac{\mathrm{C}_{1}}{\mathrm{C}_{2}}%
\langle \lambda _{1}\lambda _{2}\rangle _{cl}^{(D_{1}\&D_{2})}.  \notag
\end{eqnarray}

If $\mathrm{C}_{1}=\mathrm{C}_{2}=1$ then (\ref{35}) coincides in form with
the original Bell inequality [2-4] for the case of the perfect correlation
of the observed outcomes.

In this paper, in order to distinguish from numerous generalizations and
strengthenings of Bell's inequality, for joint experiments, classical or
quantum, we refer to an inequality\footnote{%
See [5], for a sufficient condition on the derivation of this inequality
under joint experiments on a system of any type.} between the mean values 
\begin{equation}
\left| \langle \lambda _{1}\lambda _{2}\rangle ^{(A\&D_{1})}-\langle \lambda
_{1}\lambda _{2}\rangle ^{(A\&D_{2})}\right| \leq \mathrm{C}_{1}\mathrm{C}%
_{2}-\frac{\mathrm{C}_{1}}{\mathrm{C}_{2}}\langle \lambda _{1}\lambda
_{2}\rangle ^{(D_{1}\&D_{2})},  \label{34}
\end{equation}
as the \emph{original} \emph{Bell inequality, in its perfect correlation
form.}

\end{document}